 \documentclass[final,3p,times]{elsarticle}

\usepackage{amssymb}
\usepackage{amsmath}

\usepackage{float}
\usepackage{subcaption}
\usepackage{wasysym}
\usepackage{ulem}

\usepackage{url}
\usepackage[breaklinks]{hyperref}

\usepackage{booktabs}

\usepackage{xcolor}

\journal{Applied Radiation and Isotopes}

\begin{document}

\begin{frontmatter}



\title{First investigation of $^{96}$Zr samples enriched by the gas-centrifuge method for the use in rare-decay studies}

\author[1]{D.~Arefev}
\author[2]{A.~S.~Barabash}
\author[3]{M.~De~Jesus}
\author[4]{S.~Evseev}
\author[4]{D.~Filosofov}
\author[4]{N.~Gorshkov}
\author[4,5]{V.~Kazalov}
\author[4,6]{D.~Karaivanov}
\author[4,7]{T.~Khussainov\corref{cor1}}
\author[4]{O.~Kochetov}
\author[1]{D.~Kushnarev}
\author[4,8,9]{N.~A.~Mirzayev}
\author[4,5,10]{A.~Lubashevskiy}
\author[4,5,10]{D.~Ponomarev}
\author[4,11]{A.~Rakhimov}
\author[4]{S.~Rozov}
\author[4]{K.~Shakhov}
\author[4,7]{N.~Temerbulatova}
\author[1]{D.~Timofeev}
\author[1]{A.~Ushakov}
\author[4]{S.~Vasilyev}
\author[4]{E.~Yakushev}
\author[2]{V.~Yumatov}
\author[1]{S.~Zyryanov}

\cortext[cor1]{Corresponding author. Email: khusainov@jinr.ru}

\affiliation[1]{organization={Electrochemical Plant},
city={Zelenogorsk},
postcode={663690},
country={Russia}}

\affiliation[2]{organization={NRC Kurchatov Institute, Kurchatov Complex of Fundamental Research},
city={Moscow},
postcode={117218},
country={Russia}}

\affiliation[3]{organization={Université Lyon, Université Lyon 1, CNRS/IN2P3, IP2I-Lyon},
city={Villeurbanne},
postcode={F-69622},
country={France}}

\affiliation[4]{organization={Joint Institute for Nuclear Research},
city={Dubna},
postcode={141980},
country={Russia}}

\affiliation[5]{organization={Institute for Nuclear Research of the Russian Academy of Sciences},
city={Moscow},
postcode={117312},
country={Russia}}

\affiliation[6]{organization={Institute for Nuclear Research and Nuclear Energy, Bulgarian Academy of Sciences},
city={Sofia},
postcode={BG1784},
country={Bulgaria}}

\affiliation[7]{organization={Institute of Nuclear Physics, Ministry of Energy of the Republic of Kazakhstan},
city={Almaty},
postcode={050032},
country={Kazakhstan}}

\affiliation[8]{organization={Institute of Physics, Ministry of Science and Education of Azerbaijan Republic},
city={Baku},
postcode={AZ1073},
country={Azerbaijan}}

\affiliation[9]{organization={Khazar University},
city={Baku},
postcode={AZ1096},
country={Azerbaijan}}

\affiliation[10]{organization={Lebedev Physical Institute of the Russian Academy of Sciences},
city={Moscow},
postcode={119991},
country={Russia}}

\affiliation[11]{organization={Institute of Nuclear Physics of  Uzbekistan Academy of Sciences},
city={Tashkent},
postcode={100214},
country={Uzbekistan}}

\begin{abstract}
For the first time, the isotope $^{96}$Zr has been produced by the gas-centrifuge method (enrichment is 88.28\% and full mass of $^{96}$Zr is 179.816~g). The level of radionuclide impurities of the samples and thus their suitability for rare-event search have been investigated using three low-background HPGe detectors. The most stringent limit, obtained at sea level, on the half-life of the double beta decay of $^{96}$Zr to the $0^+_1$ excited state (1148~keV) of $^{96}$Mo has been set $T_{1/2}(0\nu+2\nu)>3.9 \times 10^{19}$ yr (90\% C.L.).

\end{abstract}



\end{frontmatter}



\section{Introduction}

Two neutrino double beta decay (2$\nu$2$\beta$) is an extremely rare process allowed within the Standard Model. The main motivation for studying this process is the possibility of detecting neutrinoless double beta decay (0$\nu$2$\beta$), which is forbidden in the Standard Model. Observing such a decay would help to establish whether the neutrino is a Majorana or Dirac fermion \cite{bilenky2020neutrinos}.

Among the isotopes suitable for such studies, $^{96}$Zr stands out for several reasons. First, natural zirconium is a relatively inexpensive and readily available material, which makes it attractive for large-scale applications. Second, the Q-value of the double beta decay of $^{96}$Zr is 3356.1~keV \cite{alanssari2016single} – the third to the highest among all candidate nuclei. Such high decay energy provides a significant advantage compared to popular candidates, such as $^{76}$Ge and $^{136}$Xe. Since most of the $\gamma$-lines in the decay chains of uranium and thorium have lower energies (except for spontaneous decay) this is indeed relevant for this kind of experiments. Currently, several experiments \cite{watanabe2021development,fukuda2020zicos} are exploring double beta decay of $^{96}$Zr .

Until recently, the studies of $^{96}$Zr were limited by the difficulty in obtaining isotopically enriched material in sufficient quantities since $^{96}$Zr was obtained by electromagnetic separation. As a result, the isotope is very expensive \cite{IsotopePrices}, and only a few dozen grams of it have been produced up to now. In this work, this challenge has been overcome: for the first time, the enriched $^{96}$Zr has been produced using the gas-centrifuge method. This technological advance enables the preparation of large masses of enriched zirconium, paving the way for high-sensitivity investigations of its double beta decay.

\begin{figure}[h]
\centering
\includegraphics[width=0.9\textwidth]{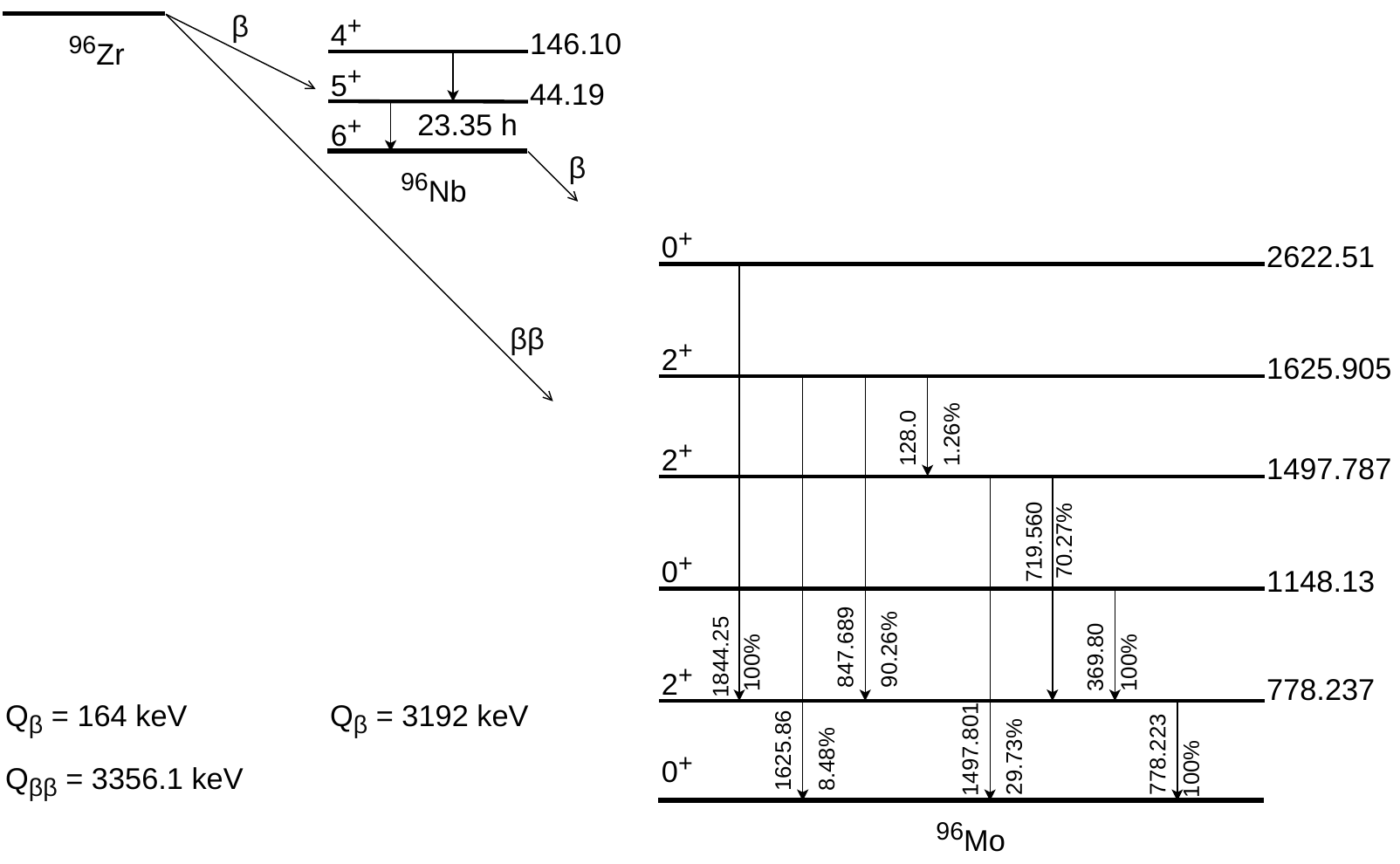}
\caption{Simplified $^{96}$Zr decay scheme \cite{abriola2008adopted}; the Q values are taken from [2].}
\label{fig:scheme}
\end{figure}

With the access to sizable and radiopure $^{96}$Zr samples, it becomes possible to explore not only the ground-state transition but also the rarer and experimentally more demanding channel of double beta decay to the excited states of $^{96}$Mo (Figure~\ref{fig:scheme}). The best current limit on the decay of $^{96}$Zr to the $0^+_1$ excited state (1148~keV) of $^{96}$Mo is $T_{1/2}(0\nu+2\nu)>3.1 \times 10^{20}$~yr \cite{finch2015search}. Detecting the $2\nu2\beta$ mode of this decay and determining its lifetime would allow further testing of theoretical models via the extraction of nuclear matrix elements.

\section{Zirconium samples}

The natural abundance of $^{96}$Zr is only 2.8\% \cite{kondev2021nubase2020}. The enrichment of zirconium in the isotope $^{96}$Zr was carried out at the Electrochemical Plant (Zelenogorsk, Russia) using a newly developed gas-centrifuge method \cite{PatentRU}. The working compound was zirconium tetrakisborohydride Zr(BH$_4$)$_4$, a volatile crystalline substance (melting point is 28.7~°C).

Natural Zr(BH$_4$)$_4$ was introduced into the separation cascade as a gas. Being the heaviest stable isotope, $^{96}$Zr concentrates as a heavy fraction. Hafnium, chemically similar to zirconium, forms a volatile hydride Hf(BH$_4$)$_4$ and also concentrates as a heavy fraction, requiring reprocessing to remove it. After repeated cascades, the material enriched in $^{96}$Zr to more than 88\% was obtained \cite{PatentRU} (see Table~\ref{tab:isotopes}) .

\begin{table}[H]
\centering
\caption{Isotopic composition of enriched zirconium.}
\begin{tabular}{cccccc}
\toprule
Isotopes & 90 & 91 & 92 & 94 & 96 \\
\midrule
Atomic fraction, \% & 0.12 & 0.04 & 0.14 & 11.42 & 88.28 \\
Mass fraction, \% & 0.11 & 0.04 & 0.14 & 11.21 & 88.50 \\
\bottomrule
\end{tabular}
\label{tab:isotopes}
\end{table}

The Zr content was provided by the manufacturer. Two forms were prepared from the enriched fraction: zirconium boride and zirconium oxide (ZrO$_2$). Zirconium boride is a mixture of various compounds of the ZrB$_x$ type and elemental boron. Considering the zirconium content in the product, the most appropriate empirical formula for the compound is ZrB$_4$. The samples were stored and transported in three identical low-background cylindrical nylon containers with outside dimensions of $ \diameter 85\times17$~mm and inside dimensions of $\diameter 79\times15.5$~mm produced of radiopure nylon by 3D printing. The same nylon was previously used for the internal part of the $\nu$GeN experiment \cite{belov2025new}. The ultralow-background N-type HPGe detector, part of the infrastructure of the EDELWEISS experiment \cite{armengaud2017performance} in the LSM underground laboratory (France), was used to measure one of the containers. No significant differences in the background spectrum were observed. The contents of the containers are described in Table~\ref{tab:samples}.

\begin{table}[H]
\centering
\caption{Zirconium sample masses.}
\begin{tabular}{ccccc}
\toprule
Container & Material & Sample mass (g) & Zr mass (g) & $^{96}$Zr mass (g) \\
\midrule
1 & Boride & 149.520 & 102.122 & 90.378 \\
2 & Oxide  & 75.865  & 56.807  & 50.274 \\
3 & Oxide  & 59.100  & 44.253  & 39.164 \\
\bottomrule
\end{tabular}
\label{tab:samples}
\end{table}

\section{Experimental strategy and setup}

As mentioned above, $^{96}$Zr has an extremely long half-life. In order to study its decay, we need low-background conditions. That means that the investigation has to be performed in an underground laboratory, for it is the only way to reduce the background associated with cosmic rays to the required level.

At the same time, there are several reasons for conducting measurements at sea level. Usually, a new setup cannot be placed in an underground laboratory immediately, and the schedule for the use of the low-background spectrometer is planned many months in advance. In addition, the radioactive purity of the sample should be evaluated beforehand to prove that it complies with the level required for conducting underground low-background measurements. Thus after the $^{96}$Zr samples were delivered to JINR, it was decided first to perform the tests at sea level and then only to transport the samples to an underground laboratory.

To study Zr samples, a new specially designed experimental setup was installed at JINR. The setup included three low-background P-type HPGe detectors with point contacts (Mirion Technologies). There were two detectors (ID 1 and 2) with the germanium crystal measuring $\diameter 70 \times70$~mm ($\sim$1.43~kg each) and one detector (ID 3) with the crystal measuring $\diameter 62\times62$~mm ($\sim$0.96~kg). The data acquisition system was similar to the one used in the $\nu$GeN experiment \cite{belov2025new}. Energy resolutions (FWHM) for the whole period of data taking at the 352 keV $\gamma$-line are equal to $1.2\pm0.2$, $1.3\pm0.2$ and $1.3\pm0.5$~keV for ID 1, 2 and 3 respectively.

\begin{figure}[H]
\centering
\includegraphics[width=0.6\textwidth]{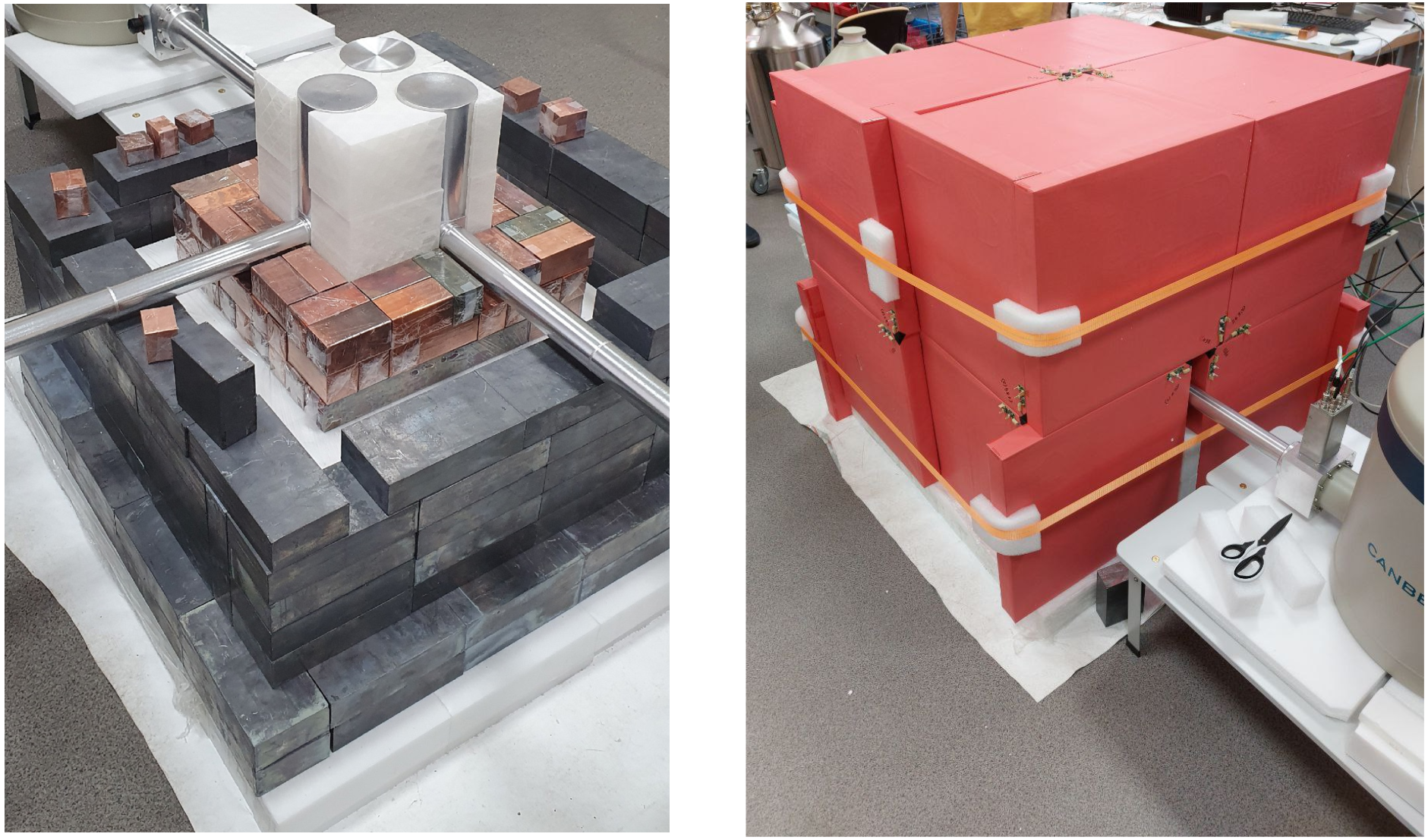}
\caption{Assembling the setup: left – assembling passive shielding around the detectors; right – final view of the setup surrounded by the muon veto.}
\label{fig:setup}
\end{figure}

The detectors were surrounded by passive and active shielding (Figure~\ref{fig:setup}). Passive shielding included 3D-printed nylon blocks to reduce air volume (and thus $^{222}$Rn quantity) followed by 10 cm of copper, 8 cm of borated polyethylene, and 10 cm of lead. Active shielding was the muon veto made of plastic-scintillator panels measuring $50\times50\times5$~cm$^{3}$. The panels were installed on all sides except the bottom.

\section{Measurements and event selection}

\begin{figure}[H]
\centering
\includegraphics[width=0.4\textwidth]{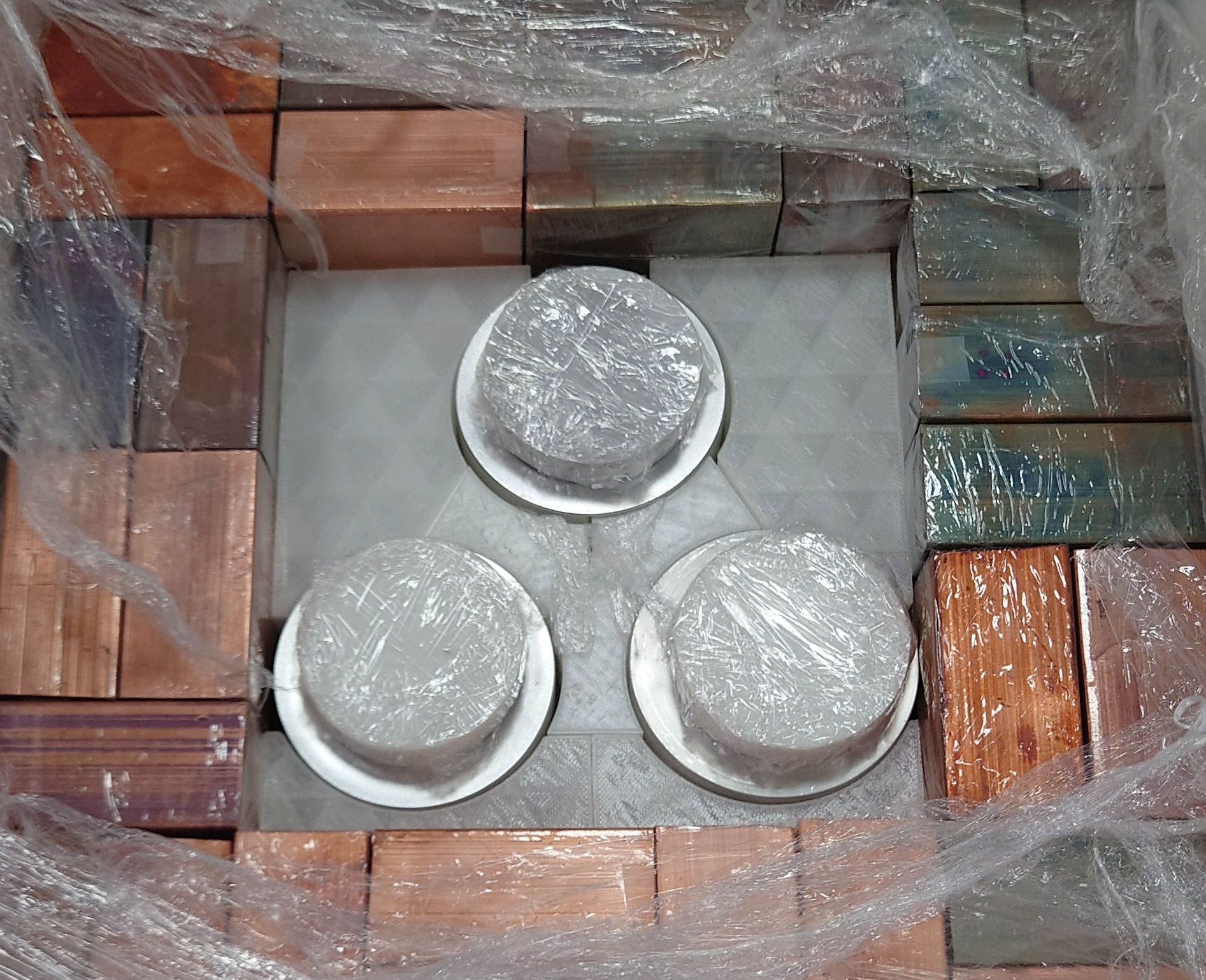}
\caption{Zirconium samples on top of detector endcaps.}
\label{fig:samples}
\end{figure}

The first stage of the study was a detailed 1179-hour measurement of the background spectrum. The zirconium samples in the above containers were put on top of the detector endcaps (Figure~\ref{fig:samples}). The spectrum with zirconium samples were measured 244 hours. 

\begin{figure}[h]
\centering
\includegraphics[width=0.8\textwidth]{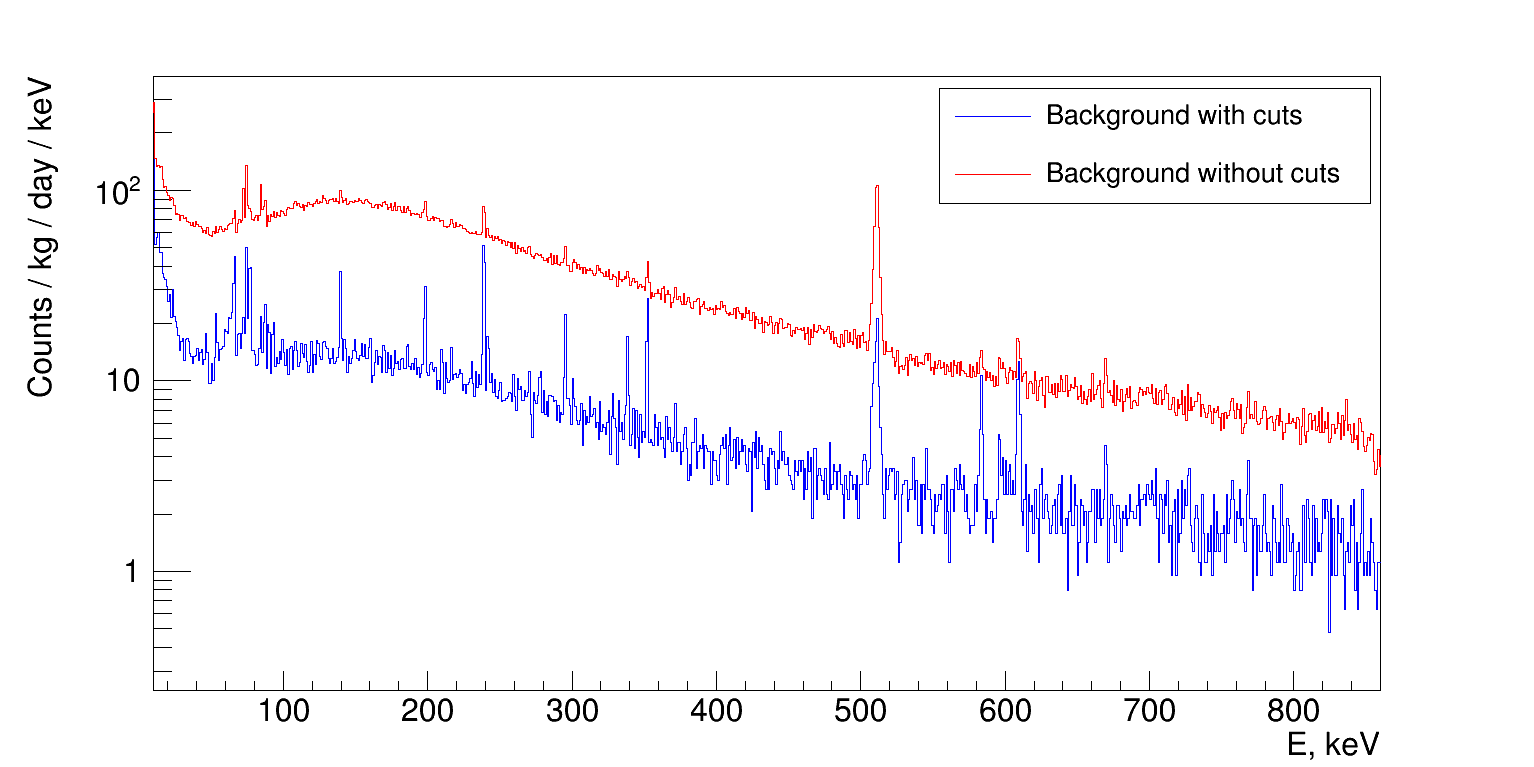}
\caption{Comparison of the background spectrum before and after the cuts.}
\label{fig:cuts}
\end{figure}

The muon veto and quality cuts reduced the overall background rate by approximately an order of magnitude (Figure~\ref{fig:cuts}). Finally, the live time of the measurement after cuts was 82\%.

\section{Monte Carlo simulations}

\begin{figure}[h]
\centering
\includegraphics[width=0.5\textwidth]{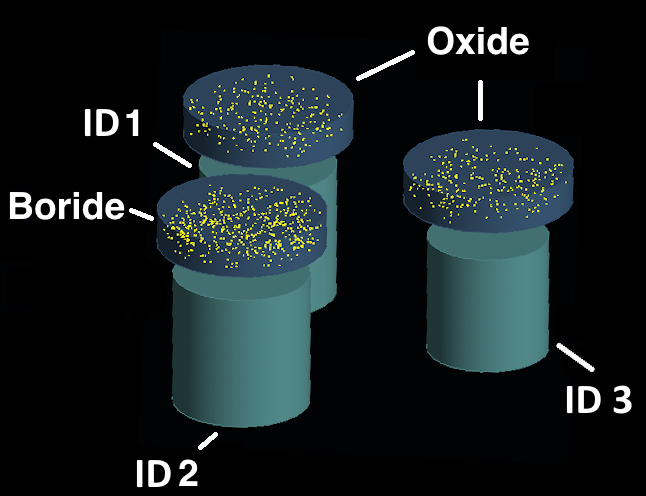}
\caption{Visualization of sample placement on top of detector endcaps with Geant4; yellow dots show the distribution of $^{96}$Zr decays in samples.}
\label{fig:MC}
\end{figure}

To evaluate the detection efficiency of the setup, the shielding, detector, and sample configurations were simulated with Geant4.10.04.p03 \cite{agostinelli2003geant4} (Figure~\ref{fig:MC}).

At that point, we assumed that zirconium boride and zirconium oxide have a similar level of radionuclide impurities. A total of $10^{7}$ decays of $^{214}$Pb (product of the $^{238}$U decay chain) and $^{208}$Tl (product of the $^{232}$Th decay chain) were generated. Each decay involved a random selection of one of the samples according to its mass and decay point, uniformly distributed in the sample. The energy deposition for the 352 ($^{214}Pb$) and 583 ($^{208}Tl$) keV $\gamma$-lines was calculated. Obtained values of the detection efficiency are presented in Table~\ref{tab:efficiencies}.

Also, $10^{7}$ decays of $^{96}$Zr to the $0^+_1$ excited state (1148~keV) of $^{96}$Mo were simulated. The decay scheme assumed the emission of two cascade $\gamma$-rays with the energies of 370 and 778~keV and the angular correlations from \cite{finch2015double}.

The detection-efficiency values shown in Table~\ref{tab:efficiencies} represent the probability that each HPGe detector registers a particular $\gamma$-ray from any of the samples. The detection probability depends both on the detector geometry and on the mass of the sample positioned above it. Therefore, detectors with heavier samples show higher efficiencies.

\begin{table}[H]
\centering
\caption{Values of the detection efficiency of the experimental setup from MC simulations (\%).}
\begin{tabular}{cccccc}
\toprule
E, keV & Decay  & ID  1 & ID 2 & ID 3 & Total\\
\midrule
352 & $^{214}$Pb$\rightarrow^{214}$Bi & 0.8 & 1.7 & 0.6 & 3.1 \\
370 & $^{96}$Zr$\rightarrow^{96}$Mo & 0.8 & 1.5 & 0.6 & 2.9 \\
583 & $^{208}$Tl$\rightarrow^{208}$Pb & 0.6 & 1.2 & 0.4 & 2.2 \\
778 & $^{96}$Zr$\rightarrow^{96}$Mo & 0.4 & 1.0 & 0.2 & 1.6 \\
\bottomrule
\end{tabular}
\label{tab:efficiencies}
\end{table}

Monte Carlo simulations of our setup were validated with a 5\% precision  using a uranium calibration source by the method described in \cite{BARABASH2026171420}.

\section{Analysis and results}

\begin{figure}[H]
\centering

\begin{subfigure}{0.65\textwidth}
    \centering
    \includegraphics[width=\linewidth]{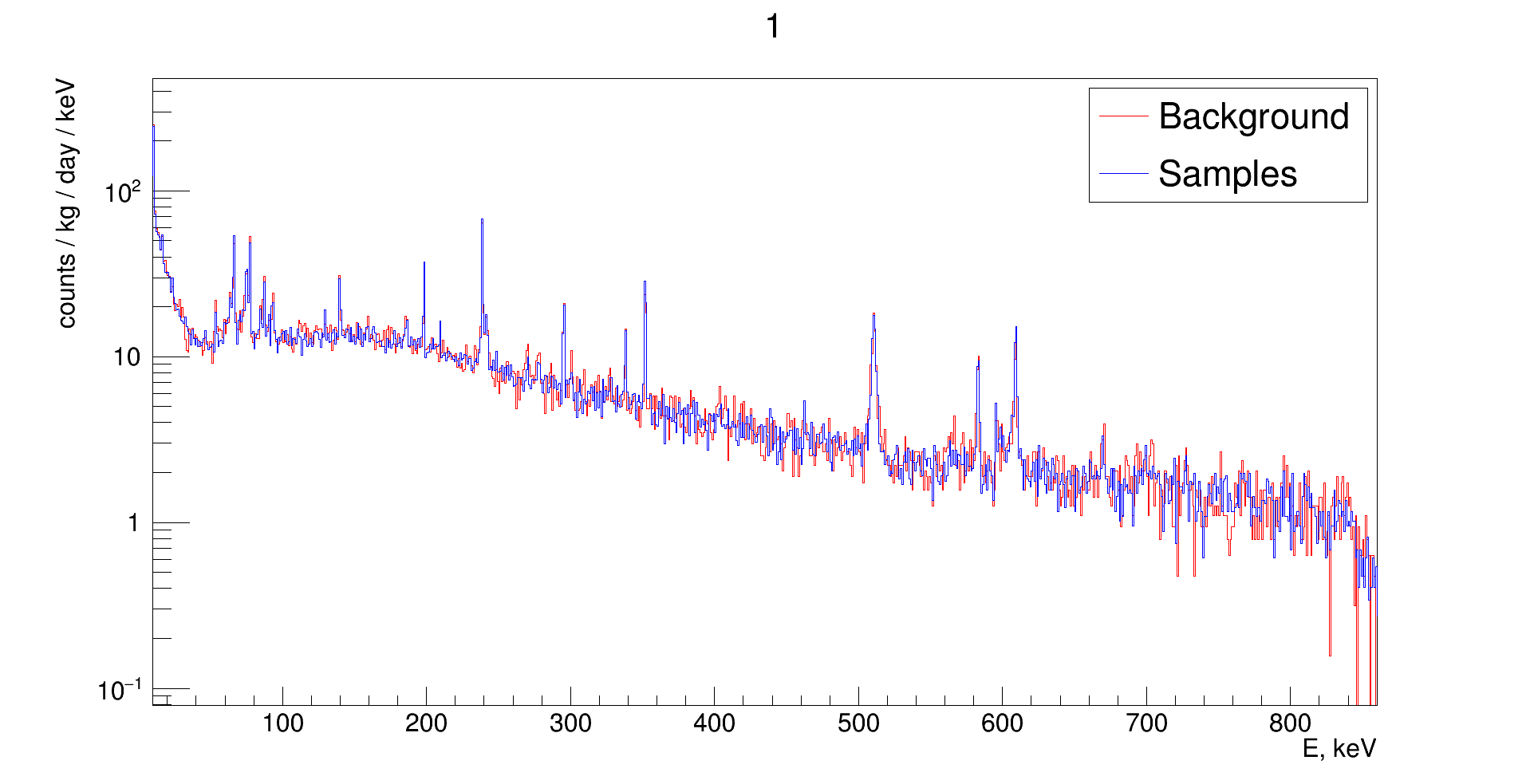}
    
\end{subfigure}

\begin{subfigure}{0.65\textwidth}
    \centering
    \includegraphics[width=\linewidth]{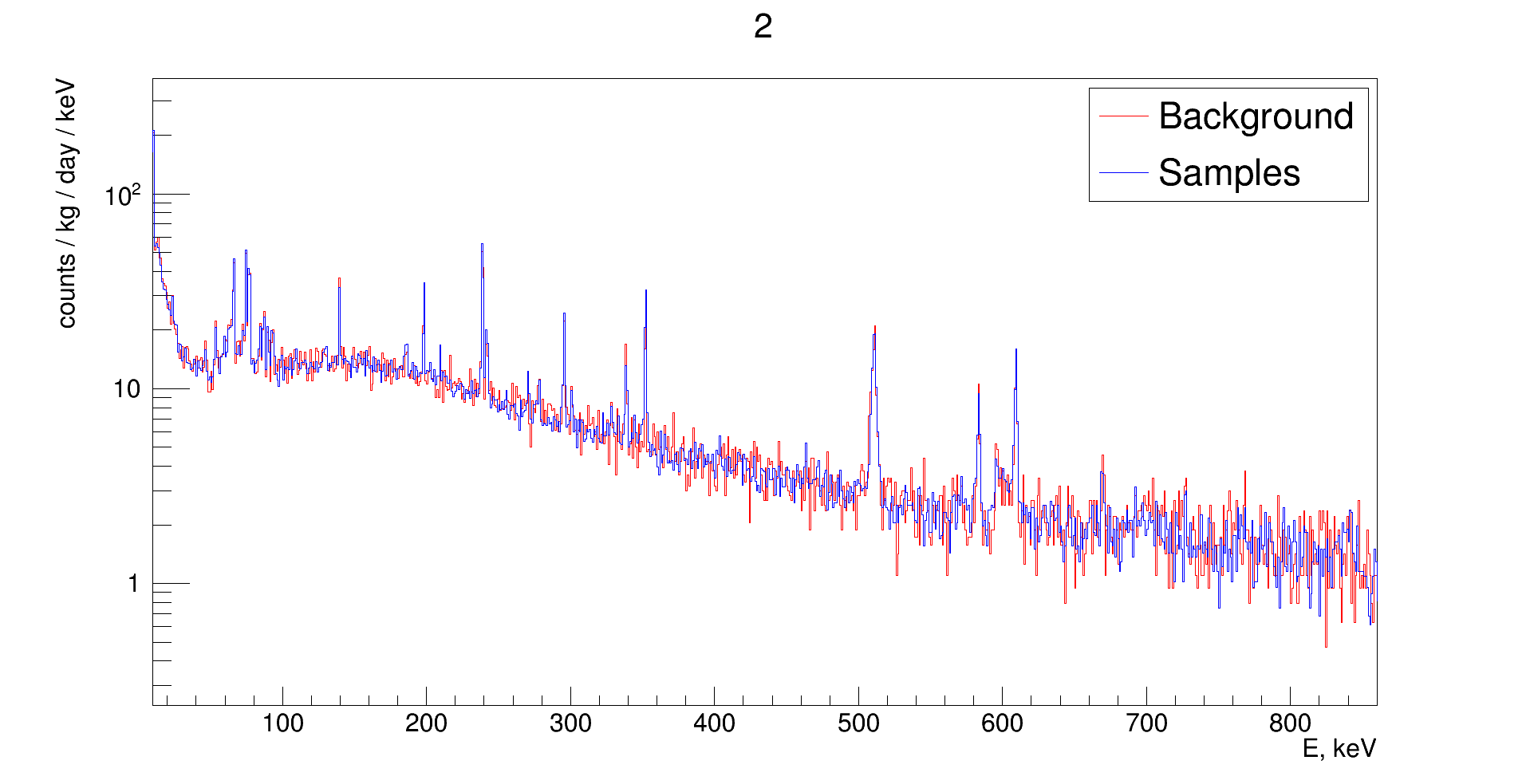}
    
\end{subfigure}

\begin{subfigure}{0.65\textwidth}
    \centering
    \includegraphics[width=\linewidth]{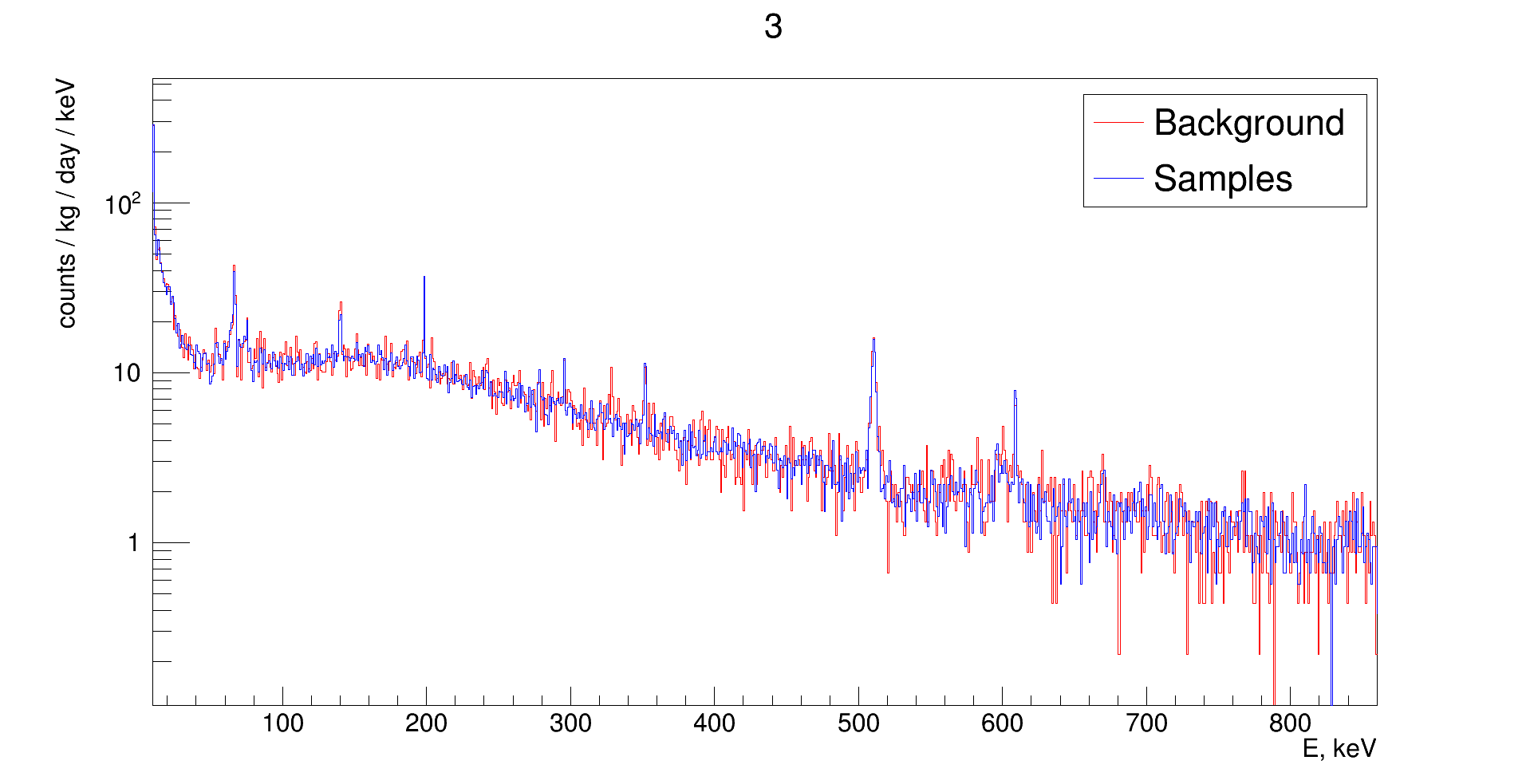}
    
\end{subfigure}

\caption{Comparison of background spectra and spectra with the samples; 1, 2 and 3 are
HPGe’s ID.}
\label{fig:comparison}
\end{figure}

The background and $\gamma$-ray experimental spectra of the samples are compared in Figure~\ref{fig:comparison}. The sample spectra exhibit no clearly visible excess over the background, neither in the full-energy peaks nor in the underlying continuum. It is the first sign of the absence of any significant contamination in our samples.

\subsection{Level of radionuclide impurities}

To analyse the uranium contamination, the 352~keV $\gamma$-line from the $^{214}$Pb decay was considered, and to analyse the thorium contamination, the 583~keV $\gamma$-line of the $^{208}$Tl decay was examined. The observed background and the sample rates for each energy are described below in Tables \ref{tab:583} and \ref{tab:352}, respectively.

\begin{table}[h]
\setlength{\tabcolsep}{8pt}
\centering
\caption{Count rates of the 583~keV line (counts/kg/day).}
\begin{tabular}{cccc}
\toprule
Detector ID & 1 & 2 & 3 \\
\midrule
Background & 17.7 $\pm$ 0.4 & 16.4 $\pm$ 0.4 & 3.6 $\pm$ 0.2 \\
Sample & 15.9 $\pm$ 1.3 & 18.0 $\pm$ 1.3 & 4.5 $\pm$ 0.7 \\
\bottomrule
\end{tabular}
\label{tab:583}
\end{table}

\begin{table}[h]
\setlength{\tabcolsep}{8pt}
\centering
\caption{Count rates of the 352~keV line (counts/kg/day).}
\begin{tabular}{cccc}
\toprule
Detector ID & 1 & 2 & 3 \\
\midrule
Background & 38.1 $\pm$ 0.6 & 40.1 $\pm$ 0.6 & 14.6 $\pm$ 0.4 \\
Sample & 43.5 $\pm$ 2.1 & 42.7 $\pm$ 2.1 & 15.2 $\pm$ 1.2 \\
\bottomrule
\end{tabular}
\label{tab:352}
\end{table}

The Feldman–Cousins statistical approach \cite{feldman1998unified} was used to quantify possible contributions from radioactive impurities. On the basis of the detection efficiency from Table~\ref{tab:efficiencies} with a total sample mass of 284.5 g, we derive an upper limit on the $^{208}$Tl activity of 31~mBq/kg with a 90\% confidence level (C.L.). The levels of radionuclide impurities in the zirconium-oxide and zirconium-boride samples are assumed to be identical.

A small difference between the sample and background counting rates at the 352~keV line may be related to radon background fluctuations. Assuming secular equilibrium of the uranium series, we interpret the result, 50~mBq/kg, as an upper limit on the $^{214}$Pb activity with a 90\% C.L.

\subsection{Double beta decay of $^{96}$Zr to the $0^+_1$ excited state of $^{96}$Mo}

There is no clearly visible excess in the spectra at energies of 370 and 778~keV. For the analysis, the regions of interest (ROIs) were defined as 370 $\pm$ 2~keV and 778 $\pm$ 2~keV (corresponding to $\sim$99 percentile). The observed background and the sample rates in the ROIs for each detector are shown in Tables \ref{tab:370} and \ref{tab:778}.

\begin{table}[h]
\setlength{\tabcolsep}{8pt}
\centering
\caption{Count rates at the 370~keV ROI (counts/kg/day).}
\begin{tabular}{cccc}
\toprule
Detector ID & 1 & 2 & 3 \\
\midrule
Background & 19.8 $\pm$ 0.4 & 18.3 $\pm$ 0.6 & 19.3 $\pm$ 0.4 \\
Sample & 17.9 $\pm$ 1.3 & 16.9 $\pm$ 1.3 & 18.4 $\pm$ 1.4 \\
\bottomrule
\end{tabular}
\label{tab:370}
\end{table}

\begin{table}[h]
\setlength{\tabcolsep}{8pt}
\centering
\caption{Count rates at the 778~keV ROI (counts/kg/day).}
\begin{tabular}{cccc}
\toprule
Detector ID & 1 & 2 & 3 \\
\midrule
Background & 4.2 $\pm$ 0.2 & 7.1 $\pm$ 0.3 & 3.1 $\pm$ 0.2 \\
Sample & 4.8 $\pm$ 0.7 & 6.5 $\pm$ 0.8 & 4.4 $\pm$ 0.7 \\
\bottomrule
\end{tabular}
\label{tab:778}
\end{table}

With the values of the detection efficiency from Table~\ref{tab:efficiencies} and a total mass of $^{96}$Zr equal to 179.816 g, the limit on the half-life of the double beta decay of $^{96}$Zr to the $0^+_1$ excited state (1148~keV) of $^{96}$Mo can be set at $T_{1/2}(0\nu+2\nu)>3.9 \times 10^{19}$~yr with a 90\% C.L. The limit is valid for both neutrinoless and two-neutrino decays and even for the decays with the Majoron emission.

The best limit for this decay to date,  $T_{1/2}(0\nu+2\nu)>3.1 \times 10^{20}$~yr \cite{finch2015search}, was obtained in the underground laboratory with a double-crystal HPGe detector operating in the coincidence mode. A sample of enriched zirconium with a total $^{96}$Zr mass of 17.9 g was investigated. The previous limits, $T_{1/2}(0\nu+2\nu)>3.3 \times 10^{19}$~yr \cite{arpesella1994search} and $T_{1/2}(0\nu+2\nu)>6.8 \times 10^{19}$~yr \cite{barabash1996investigation}, were obtained with different single-crystal HPGe detectors in the underground laboratories LNGS (Italy) and LSM (France) by examining the sample of enriched zirconium with a total $^{96}$Zr mass of 8.16 g and the 57\% enrichment. It is clear that measurements using our 179.816 g of $^{96}$Zr in an underground laboratory will significantly increase the sensitivity of the experiment, which gives a hope to detect the $2\beta(2\nu)$ decay of $^{96}$Zr into the $0^+_1$ excited state of $^{96}$Mo.

\section{Conclusion}

For the first time, zirconium enriched in the isotope $^{96}$Zr has been produced by the gas-centrifuge method, opening the possibility of obtaining large masses of material for next-generation studies of double beta decay. $^{96}$Zr has the highest energy among all candidates for double beta decay produced by the gas-centrifuge method so far and thus is one of the most promising candidate for searching for neutrinoless double beta decay. The combined amount of $^{96}$Zr in our samples (179.816~g) exceeds that in each of the previous studies \cite{finch2015search, arpesella1994search, barabash1996investigation} by more than an order of magnitude. 

The measurements performed with three HPGe detectors yielded the most stringent sea-level limit on the half-life of the double beta decay of $^{96}$Zr to the $0^+_1$ excited state (1148~keV) of $^{96}$Mo corresponding to $T_{1/2}(0\nu+2\nu)>3.9 \times 10^{19}$ yr. The radionuclide impurities of the enriched samples considering  $^{214}$Pb and  $^{208}$Tl have been found to be below the level of 50~mBq/kg and 31~mBq/kg, respectively, demonstrating their suitability for rare-event experiments.

These results establish a solid motivation to use the obtained samples in an underground laboratory to perform measurements with a substantially improved sensitivity.

\section{Acknowledgments}
This work is supported by the State Project "Science" by the Ministry of Science and Higher Education of the Russian Federation under Contract 075-15-2024-541.




\begin{thebibliography}{00}

\bibitem{bilenky2020neutrinos}
S.~M.~Bilenky,
Neutrinos: Majorana or Dirac?,
\textit{Universe} \textbf{6} (2020) 134.

\bibitem{alanssari2016single}
M.~Alanssari, D.~Frekers, T.~Eronen, L.~Canete, J.~Dilling, M.~Haaranen,
J.~Hakala, M.~Holl, M.~Je{\v{s}}kovsk{\`y}, A.~Jokinen \textit{et al.},
Single and double beta-decay Q values among the triplet $^{96}$Zr, $^{96}$Nb, and $^{96}$Mo,
\textit{Phys. Rev. Lett.} \textbf{116} (2016) 072501.

\bibitem{watanabe2021development}
A.~Watanabe, A.~Magi, M.~Koshimizu, A.~Sato, Y.~Fujimoto, K.~Asai,
Development of Zr-loaded green-emitting liquid scintillator for detection of neutrinoless double beta decay,
\textit{Sens. Mater.} \textbf{33} (2021) 2251--2261.

\bibitem{fukuda2020zicos}
Y.~Fukuda, S.~Moriyama, K.~Hiraide, I.~Ogawa, T.~Gunji, R.~Hayami,
S.~Tsukadaw, S.~Kurosawa,
ZICOS—Neutrinoless double beta decay experiment using $^{96}$Zr with an organic liquid scintillator,
\textit{J. Phys.: Conf. Ser.} \textbf{1468} (2020) 012139.


\bibitem{IsotopePrices}
{Institut f\"ur Seltene Erden},
Prices for stable isotopes,
Available at: \url{https://ritverc.com/en/products/reference-and-check-sources-and-solutions/gamma-sources/point-sources} (accessed 6 Oct 2025).

\bibitem{finch2015search}
S.~W.~Finch, W.~Tornow,
Search for two-neutrino double-$\beta$ decay of $^{96}$Zr to excited states of $^{96}$Mo,
\textit{Phys. Rev. C} \textbf{92} (2015) 045501.

\bibitem{abriola2008adopted}
D.~Abriola, A.A.~Sonzogni,
Nuclear Data Sheets for A = 96,
\textit{Nuclear Data Sheets} \textbf{109} (2008) 2501.

\bibitem{kondev2021nubase2020}
F.~G.~Kondev, M.~Wang, W.~J.~Huang, S.~Naimi, G.~Audi,
The NUBASE2020 evaluation of nuclear physics properties,
\textit{Chin. Phys. C} \textbf{45} (2021) 030001.

\bibitem{PatentRU}
Method for separating zirconium isotopes,
Russian Federation Patent RU2794182 (2023).

\bibitem{belov2025new}
V.~Belov, A.~Bystryakov, M.~Danilov, S.~Evseev, M.~Fomina,
G.~Ignatov, S.~Kazartsev, J.~Khushvaktov, T.~Khussainov,
A.~Konovalov \textit{et al.},
New constraints on coherent elastic neutrino--nucleus scattering by the $\nu$GeN experiment,
\textit{Chin. Phys. C} \textbf{49} (2025) 053004.

\bibitem{armengaud2017performance}
E.~Armengaud, Q.~Arnaud, C.~Augier, A.~Beno{\^\i}t, L.~Berg{\'e},
T.~Bergmann, J.~Billard, T.~De~Boissi{\`e}re, G.~Bres,
A.~Broniatowski \textit{et al.},
Performance of the EDELWEISS-III experiment for direct dark matter searches,
\textit{J. Instrum.} \textbf{12} (2017) P08010.

\bibitem{agostinelli2003geant4}
S.~Agostinelli, J.~Allison, K.~Amako, J.~Apostolakis, H.~Araujo,
P.~Arce, M.~Asai, D.~Axen, S.~Banerjee, G.~Barrand \textit{et al.},
Geant4—a simulation toolkit,
\textit{Nucl. Instrum. Meth. A} \textbf{506} (2003) 250--303.

\bibitem{finch2015double}
S.~W.~Finch,
\textit{Double Beta Decay Studies in $^{96}$Zr},
Ph.D. thesis, Duke University (2015).

\bibitem{BARABASH2026171420}
A.S.~Barabash, S.~Evseev, D.~Filosofov, V.~Kazalov, T.~Khussainov,
A.~Lubashevskiy, N.~D.~Mokhine, D.~Ponomarev, S.~Rozov,
S.~Vasilyev \textit{et al.},
Application of a high-precision distributed uranium source for determining the effective mass and volume of a HPGe detector,
\textit{Nucl. Instrum. Meth. A} \textbf{1087} (2026) 171420.

\bibitem{feldman1998unified}
G.~J.~Feldman, R.~D.~Cousins,
Unified approach to the classical statistical analysis of small signals,
\textit{Phys. Rev. D} \textbf{57} (1998) 3873-3889.

\bibitem{arpesella1994search}
C.~Arpesella, A.~S.~Barabash, E.~Bellotti, C.~Brofferio,
E.~Fiorini, P.~P.~Sverzellati, V.~I.~Umatov,
Search for $\beta\beta$ decay of $^{96}$Zr and $^{150}$Nd to excited states of $^{96}$Mo and $^{150}$Sm,
\textit{Europhys. Lett.} \textbf{27} (1994) 29-34.

\bibitem{barabash1996investigation}
A.S. Barabash, R. Gurriaran, F. Hubert, Ph. Hubert. J.L. Reyss, J. Suhonen, V.I. Umatov,
Investigation of the $\beta\beta$ decay of $^{96}$Zr to excited states in $^{96}$Mo,
\textit{J. Phys. G: Nucl. Part. Phys.} \textbf{22} (1996) 487-496.

\end{thebibliography}

\end{document}